\documentclass{article}

\PassOptionsToPackage{numbers, compress}{natbib}

\usepackage{aimc2022}

\usepackage[utf8]{inputenc} 
\usepackage[T1]{fontenc}    
\usepackage{hyperref}       
\usepackage{url}            
\usepackage{booktabs}       
\usepackage{amsfonts}       
\usepackage{nicefrac}       
\usepackage{microtype}      
\usepackage{graphicx}       
\usepackage{here}

\title{MR4MR: Mixed Reality for Melody Reincarnation}

\author{
    Atsuya Kobayashi \\
    Graduate School of Media and Governance\\
    Keio University\\
    5322 Endo, Fujisawa City, Kanagawa Japan \\
    \texttt{atsuya@sfc.keio.ac.jp}
    \AND
    Ryogo Ishino \\
    Faculty of Environment and Information Studies \\
    Keio University\\
    5322 Endo, Fujisawa City, Kanagawa Japan \\
    \texttt{t18063ri@sfc.keio.ac.jp}
    \And
    Ryuku Nobusue \\
    Graduate School of Media and Governance \\
    Keio University\\
    5322 Endo, Fujisawa City, Kanagawa Japan \\
    \texttt{ryuku@sfc.keio.ac.jp}
    \And
    Takumi Inoue \\
    Faculty of Policy Management \\
    Keio University\\
    5322 Endo, Fujisawa City, Kanagawa Japan \\
    \texttt{s19086ti@sfc.keio.ac.jp} 
    \And
    Keisuke Okazaki \\
    Faculty of Environment and Information Studies \\
    Keio University \\
    5322 Endo, Fujisawa City, Kanagawa Japan \\
    \texttt{t19152ko@sfc.keio.ac.jp} 
    \And
    Shoma Sawa \\
    Faculty of Environment and Information Studies \\
    Keio University \\
    5322 Endo, Fujisawa City, Kanagawa Japan \\
    \texttt{t19563ss@sfc.keio.ac.jp}
    \And
    Nao Tokui \\
    Graduate School of Media and Governance \\
    Keio University \\
    5322 Endo, Fujisawa City, Kanagawa Japan \\
    \texttt{tokui@sfc.keio.ac.jp}
}

\begin{document}

\maketitle

\begin{abstract}
  There is a long history of an effort made to explore musical elements with the entities and spaces around us, such as musique concrète and ambient music. In the context of computer music and digital art, interactive experiences that concentrate on the surrounding objects and physical spaces have also been designed. In recent years, with the development and popularization of devices, an increasing number of works have been designed in Extended Reality to create such musical experiences. In this paper, we describe MR4MR, a sound installation work that allows users to experience melodies produced from interactions with their surrounding space in the context of Mixed Reality (MR). Using HoloLens, an MR head-mounted display, users can bump virtual objects that emit sound against real objects in their surroundings. Then, by continuously creating a melody following the sound made by the object and re-generating randomly and gradually changing melody using music generation machine learning models, users can feel their ambient melody "reincarnating".
\end{abstract}

\section{Introduction}

In computer music and digital artworks, many interactive experiences that focus on the objects and physical space around us have been designed so far. In the past few years, XR (Extended Reality) environments have also become easily accessible, and this has led to the development of the research field of Musical XR \cite{music_in_xr}. Mixed Reality (MR) with head-mounted displays can be used for spatial music experiences since it can provide the user an immersive experience like Virtual Reality (VR), while interacting with the real space like Augmented Reality (AR).

Exploring musical elements with the entities and space around us has been challenged for a long time, such as musique concrète and ambient music. In the context of computer music and digital art, interactive experiences that concentrate on the spatial environment: surrounding objects and physical spaces have also been designed. In the context of MR, which can handle human physical movements in virtual space, various systems of musical experience have been proposed for both listening and composing music. As for the design of music listening, an interface for surround music listening is proposed, that allows the user to design a free surround space in an MR environment by freely arranging sound sources in the real space \cite{placing_music_in_space}. And a musical mobile application is proposed that added a spatial dimension to a music mobile application to allow users to listen to a song as if it were played live \cite{spatial_music_listening_experience}. On the other hand, in the context of composition and performance support, Music Everywhere \cite{ar_piano_impro} proposes an interface to support performance practice from a virtual display on top of the real keyboard in MR. MiXR \cite{mixr} is also an MR-based system that provides musical instrument players with controllable musical score sheets presentation. Besides, in the field of sound art, Sonic Sculpture \cite{sonic_sculpture} proposes to extend the audience’s approach to sound in the experience of sound sculpture by wearing a head-mounted display. Brian Eno has designed an experience that creates sound art from spatial interactions \cite{bloom_2018}. This includes a synthesizer interface that creates ambient music from curves drawn by the user, and a system that allows multiple people to experience audiovisual effects from mid-air drawing line interaction \cite{gen_music_ar_drawing}. Multimodal connectoR \cite{komatsubara_ohta_ohashi_sonobe_nakagawa_2019} is an application that users can experience sounds that match the collision position of virtual objects in the MR space. A library OSC-XR \cite{osc_xr} has been proposed for the design of music interfaces on XR as described above. There are also actual contents of music experiences with XR provided as services: such as Drops: Rhythm Garden\footnote{Drops: Rhythm garden \url{https://store.steampowered.com/app/864960/Drops_Rhythm_Garden/}}, which is a game to enjoy the sound of objects' movements in VR space, and Fields\footnote{Fields, Planeta.Cc  \url{https://fields.planeta.cc/}}, which offers the experience of creating music from the surrounding environment. Some companies design musical experiences and tools for the XR environment\footnote{Emotionwave XR \url{https://emotionwave.com/xr/}}\footnote{PatchXR - Sound of the Metaverse \url{https://patchxr.com/}}. XR is very useful as a platform for designing new musical experiences and will be further developed in the future.

In the area of musical artificial intelligence, fairly sophisticated interpretation and generation have become possible and are beginning to be used for recommendation and composition support respectively; Google Magenta offers Magenta Studio\footnote{Magenta Studio. Magenta. \url{https://magenta.tensorflow.org/studio/}}as a tool, which applies MusicVAE model \cite{music_vae} for music production, and proposes an attempt to extend a number of musical experiences as interactive demos. Adapting automatic composition algorithms to the design of new musical interactions could provide novel musical experiences. As an example of an application to audibility, \cite{smart_city_sonification} applies MusicVAE to the conversion of data flows into sound in Dublin's Smart City, proposing a way to utilize music generation models beyond composition support. In addition, Malakai \cite{malakai}, a tool that covers the process of remixing, uses MusicVAE to show the flow of music composition in real-time to accompany an interactive experience, showing the potential of applying artificial intelligence to musical experiences outside the context of normal DAW composition. A review on artificial intelligence and XR \cite{review_ai_and_xr} analyzes machine learning approaches on many XRs, but no research cases on music are presented. As mentioned above, music artificial intelligence can be a method to create various sound experiences from a wide variety of interactions, so we design an experience that generates Background Music from spatial interactions.


In this paper, we propose the interactive system MR4MR, which realizes the novel experience of listening to the music generated from the user's interaction with surrounding physical spaces. With MR4MR, users can give birth to a new melody by touching virtual objects floating in the MR space, and listening to the gradual change, which we called "Reincarnation", of the melody once created by the user. In the following sections, we describe the detail design / implementation of the application, evaluation of the experience through feedback in installation exhibition, and discussion about the technical limit and future works. A demo video is available at \url{https://youtu.be/xPMbqHUlyyU}

\section{Usage and System Design}

\begin{figure}[h]
  \centering
  \includegraphics[width=1\linewidth]{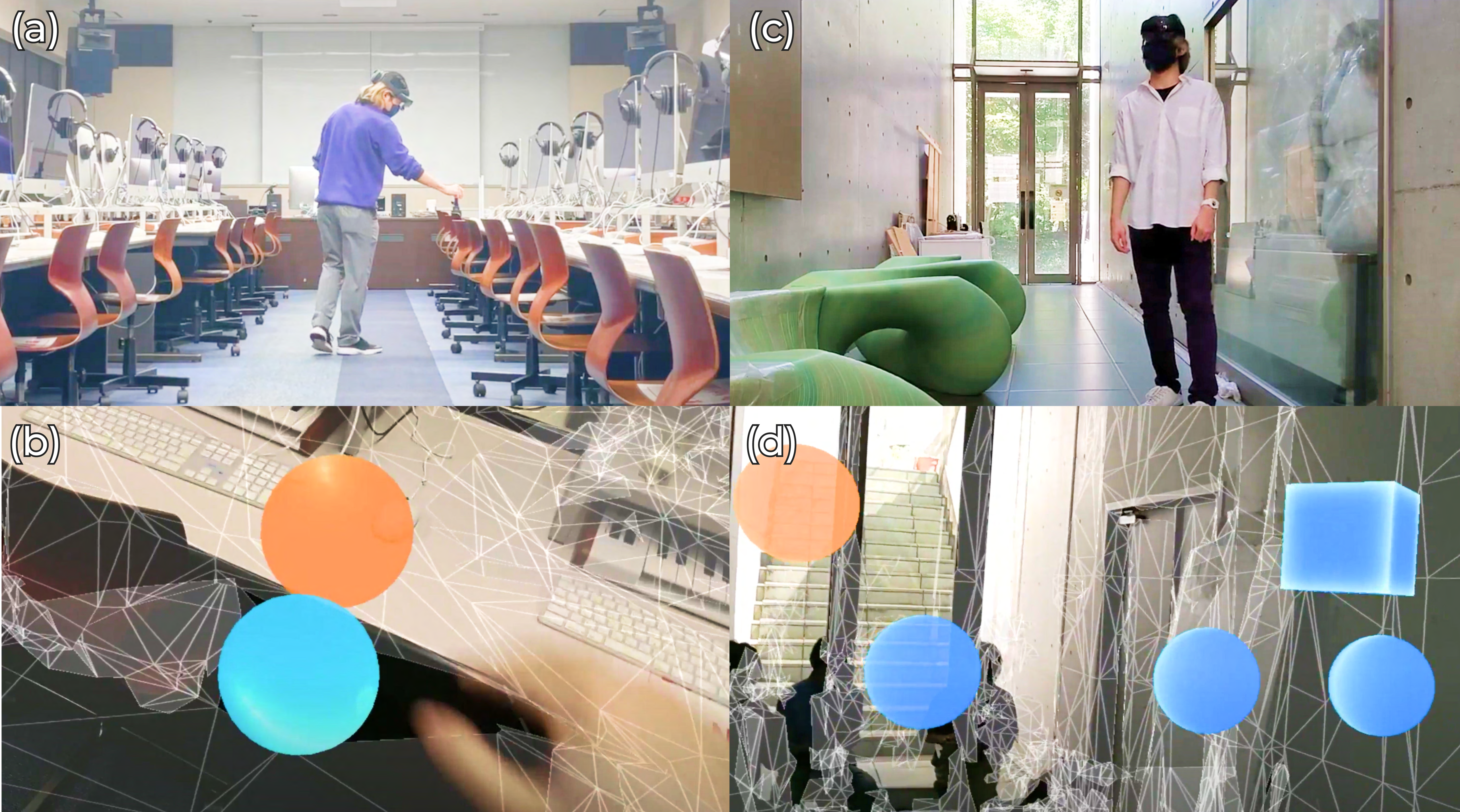}
  \caption{Fig (a) shows the MR4MR being experienced in the computer room, and the view is shown in Fig (b). Fig (c) shows a corridor in a building, and the view from the player is as shown in Fig (d).}
  \label{experience}
\end{figure}

To experience MR4MR, the user puts on the HoloLens head-mounted display and wireless headphones and launches the application on the MR field. Three kinds of 3D objects are spawned at the starting point, and the user can pick up the balls, move them, or hit them to make them glide. When these objects collide with walls or objects in real space, such as chairs, tables, laptops, etc., they generate a collision sound and bounce back. We defined gravity as well in the virtual space, and once the user touches the object, it falls to the floor and starts bouncing. This force of gravity can be modified in the settings and is also changed when a particular object collides (Figure \ref{experience}).

\begin{figure}[H]
  \centering
  \includegraphics[width=1\linewidth]{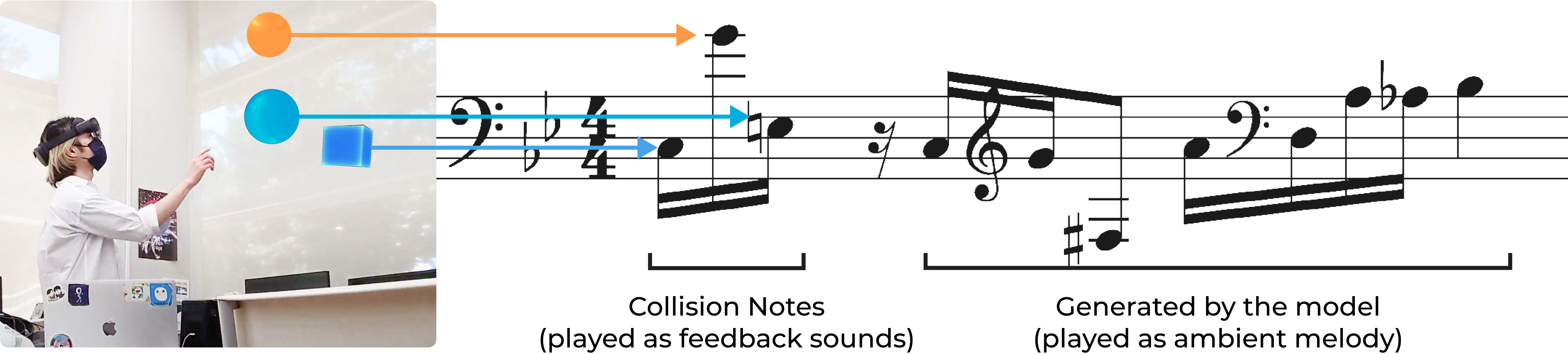}
  \caption{Generative models create the sequence of notes that follows the input base melody.}
  \label{generation-flow}
\end{figure}

Suppose some collision sounds get played more than a certain number of times within a given period. Then, we treat them as a beginning of a base melody piece and feed them into generative melody models to generate the following melody (Figure \ref{generation-flow}). Once the generated melody plays to the end, it will loop, but the melody will change slightly every few bars by the method described later. In addition, the scale of the melody changes depending on the ambiance of the place where the object hits, so users can enjoy the changing background tune as they reposition from one spot to another, such as a dark place, a bright room, or a room with colorful and vividly colored walls.

This system consists of 4 modules. 

\begin{itemize}
    \item Melody sound module integrated with Ableton Live 11\footnote{Ableton Live 11, \url{https://www.ableton.com/en/live/}} and Max for Live\footnote{Max for Live, \url{https://www.ableton.com/en/live/max-for-live/}}
    \item Python server for handling collision information as a note and managing music generation models
    \item MR interface module run on Microsoft HoloLens\footnote{Microsoft HoloLens, \url{https://www.microsoft.com/en-us/hololens}}
    \item Controller for retrieving visual information of the field of vision in MR space.
\end{itemize}

\begin{figure}[H]
  \centering
  \includegraphics[width=1\linewidth]{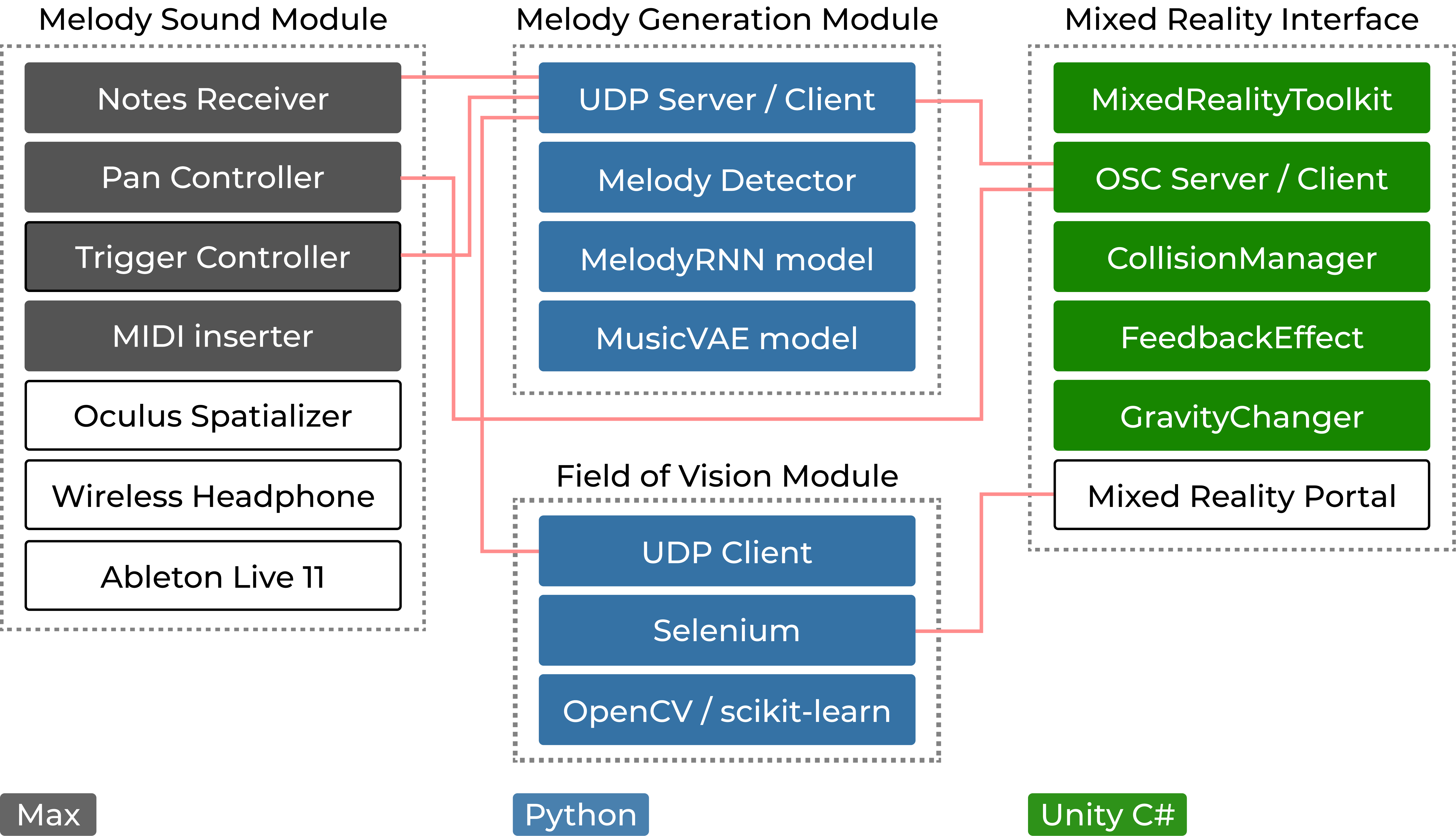}
  \caption{System Architecture: MR4MR application consists of 4 modules developed with Max, Python, and Unity. Each module has sub-modules separated to handle specific functions and sub-modules connected via OSC to each other to communicate with other modules.}
  \label{system-architecture}
\end{figure}

These modules are linked via OSC (Open Sound Control\footnote{Open Sound Control, \url{https://opensoundcontrol.stanford.edu/}}) (Figure \ref{system-architecture}). Modules other than the MR interface can be run on both a single computer and multiple computers connected to the same network.

\subsection{Mixed Reality Interface}

Using Microsoft's Mixed Reality Toolkit\footnote{Mixed Reality Toolkit \url{https://github.com/microsoft/MixedRealityToolkit-Unity}}, which delivers a set of features for inventing interactive scenes with HoloLens, we designed 3 types of interactive virtual 3D objects in Unity\footnote{Unity Real-Time Development Platform | 3D, 2D VR AR Engine, \url{https://unity.com/}}. Each of the 3 types has a distinguishable collision sound, and one of them randomly changes the gravity setting in the MR environment when it collides. Information of type, velocity, and coordinates of each object are sent to the melody generation module when the object collides with a 3D mesh surface generated on a wall or the surface of an object in real space.

Objects in real space, such as walls and desks, are also recognized in real-time by the field view camera mounted on the HoloLens, and a 3D mesh is generated along with them, so virtual objects can bounce and fly when they collide with real objects. Consequently, if there are many objects in the space where the user plays MR4MR, the number of sounds will increase, and following melodies will be generated incessantly.

\subsection{Melody Sound Module}

Users can design the atmosphere of collision sounds since they can edit the mapping of timbre and each object in Ableton Live if they want. A sound sample of a bell, xylophone, or glockenspiel is assigned by default. Collision sound has a pitch based on the height of the collision location and a volume according to the collision velocity. Using the Oculus Spatializer VST plugin\footnote{Oculus VST Spatializer \url{https://developer.oculus.com/documentation/native/audio-osp-vst/}}, collision sound feedback is processed to surround sound to enrich the fidelity and enable the user to hear collision sounds from a precise location. 

As a implementation of this function, every received object information consisting of the coordinates of the user and the coordinates and speed of the object is converted into a MIDI note and Control Change value. The module consist of a Max patch and a Max for Live device respectively, each with the sub-modules described in Fig.\ref{system-architecture}. The patch has sub-modules of Note Receiver and Pan Controller for receiving the information via OSC and send spatial information as MIDI Control Change to Oculus Spatializer Plugin. Max for Live device has sub-modules of MIDI inserter and Trigger Controller to receive generated melody MIDI note sequence via OSC and create MIDI clip on Ableton Live session view via Live API.

\subsection{Melody Generation Module}

In MR4MR, the user not only hears the collision sounds but also listens to the tune of automatically generated melody that follows the tune of collision sounds. When more than a certain number of collision sounds occur in a certain period, the system regards the collision sound sequence as a base melody component and generates the following part of melody with a trained MelodyRNN \cite{repo_melody_rnn} model. Generated melody piece is played in the loop with an instrument assigned in Ableton Live. As the generated melody is played back, notes information is sent back to the HoloLens for visual feedback. While playing the generated melody, a blue wave-like graphic is projected into MR space to show the user from which height collision the melody was generated.

If the generated melody is looped more than a certain number of times, a slightly modified melody is generated and overwrites the currently looped melody. The new melody is reconstructed from the sum of random noise and latent vector got by encoding the previous melody into a public trained 2-bar MusicVAE \cite{repo_music_vae} model. In this model, a bar of monophonic melodies in sequences of 16th note events are used. Approximately 1.5 million MIDI files collected from web were used to train the model. If there are fewer objects in the physical space of the MR4MR experience or fewer interactions with virtual objects, the user can enjoy the intermittent, gradually changing flow of a spawned melody. The continuous modification of the loop melody is achieved by adding a random noise to the latent vector $z$ encoded from previous loop to generate the next loop in every two bars. Random noise is a vector with the same size as $z$ and each element sampled from a uniform distribution of min 0 and max $1.0^{-4}$.

\subsection{Field of Vision Module}

\begin{figure}[h]
  \centering
  \includegraphics[width=1\linewidth]{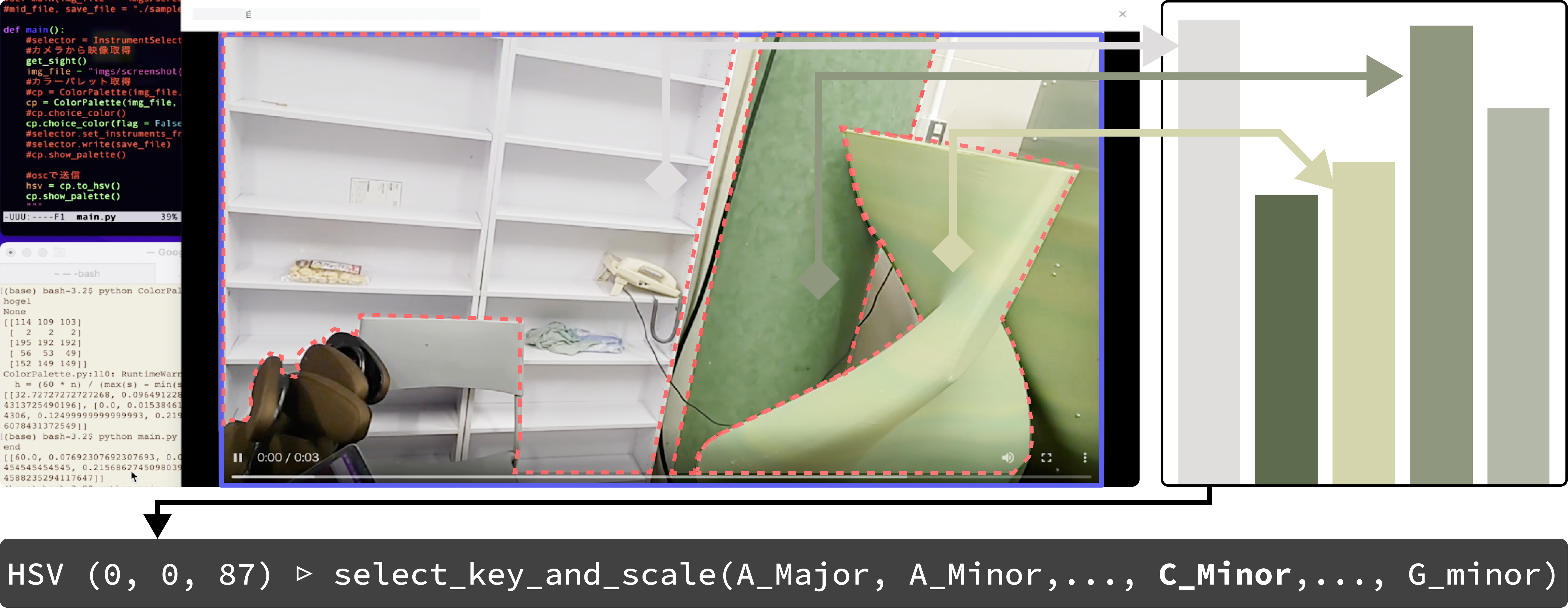}
  \caption{Getting HSV color data from a screenshot image of the field of view camera mounted on HoloLens.}
  \label{scene-tracking}
\end{figure}

By extracting the color information, this system converts the generated melody scale to match the mood of the space. The image of the HoloLens' field of view is captured once every few seconds via web interface pre-built in HoloLens. Each pixel of the acquired image is clustered by RGB value with the k-means algorithm (Figure \ref{scene-tracking}). The color value that makes up the majority of the screen area is sent to the music generation module. Each key and scale will have its own distinctive emotional image, and each type of color and brightness will have its own associated impression, then the color information can be converted into keys and scales. 

Alexander Scriabin proposed to correspond mappings of color and tone. Project Scriabin \cite{project_scriabin} is an interactive system of synesthetic experience. We developed a key and scale converter on the basis of the table (Table \ref{color-scale-mapping}), which was formulated with reference to Scriabin's color to scale mapping. In the previous work \cite{mardirossian-2007-visualizing}, the value of Hue was defined as a single scalar value for each tonic of the key. In this work, however, each boundary between colors was set to the value that is equidistant from the adjacent colors. In addition, based on the general impression that major tunes are brighter and minor tunes are darker, we set a threshold of Saturation and Value between major and minor scales.

\begin{table}[ht]
    \caption{Color Range to Scale Mappings}
    \label{color-scale-mapping}
    \centering
    \begin{tabular}{llllll}
        \toprule
        Typical Color & Hue & Saturation & Value & Tonic & Scale \\
        \midrule
        Red & [331, 360] & [.0, .6] & [.7, 1.0] & C & Major \\
        Maroon Color & [0, 19] & [.0, .6] & [.0, .7) & C & Minor \\
        Light Orange & [20, 49] & [.7, 1.0] & [.8, 1.0] & G & Major \\
        Mustard Yellow & [20, 49] & [.3, 1.0] & [.5, .8) & G & Minor \\
        Canary Yellow & [50, 90] & [.0, .7) & [.8, 1.0] & D & Major \\
        Olive Color & [50, 90] & [.4, 1.0] & [.2, .8) & D & Minor \\
        Neon Green & [91, 140] & [.2, 1.0] & [.8, 1.0] & A & Major \\
        Dark Green & [91, 140] & [.5, 1.0] & [.1, .8) & A & Minor \\
        Aquamarine Color & [141, 200] & [.0, .3) & [.8, 1.0] & E & Major \\
        Teal Color & [141, 200] & [.2, 1.0] & [.2, .7] & E & Minor \\
        Blue & [201, 248] & [.2, 1.0] & [.6, 1.0] & B & Major \\
        Navy Blue & [201, 248] & [.2, .7] & [.0, .6) & B & Minor \\
        Blue Violet & [249, 265] & [.6, 1.0] & [.7, 1.0] & F\# & Major \\
        Indigo Purple & [249, 265] & [.0, .7] & [.0, .7) & F\# & Minor \\
        Heliotrope Purple & [266, 277] & [.0, .5) & [.5, 1.0] & C\# & Major \\
        Koki Murasaki & [266, 283] & [.1, 1.0] & [0., .5) & C\# & Minor \\
        Magenta Color & [284, 310] & [.4, 1.0] & [.6, 1.0] & G\# & Major \\
        Byzantium Color & [284, 310] & [.1, 1.0] & [.2, 6) & G\# & Minor \\
        Azalea Pink & [311, 330] & [.6, 1.0] & [.2, 1.0] & A\# & Major \\
        English Violet & [311, 330] & [.1, .6) & [.1, .4] & A\# & Minor \\
        Flamingo Pink & [331, 350] & [.2, .8] & [.6, 1.0] & F & Major \\
        Tyrian Purple & [331, 350] & [.2, 1.0] & [.2, .7] & F & Minor \\
        \bottomrule
    \end{tabular}
\end{table}

\section{Evaluation}

\begin{table}[H]
  \caption{Results of 5-point Likert scale questions}
  \label{eval-result}
  \centering
  \begin{tabular}{ll}
    \toprule
    Question & Result (mean) \\
    \midrule
    Was the experience interesting? & 4.67±0.73 \\
    How pleasant was your experience? & 4.22±0.97 \\
    Was the experience is unique? & 4.56±0.93 \\
    Do you think the experience was a musical experience? & 4.22±0.80 \\
    \bottomrule
  \end{tabular}
\end{table}

\begin{table}[H]
  \caption{Feedback Comments from Participants}
  \label{eval-comments}
  \centering
  \begin{tabular}{cl}
    \toprule
    Participant No. & Feedback Comment \\
    \midrule
    4 & The experience with Hololens was amazing! \\
    8 & The fusion of MR and music-generating AI was interesting! \\
    9 & I thought it would be fun to make the music more tangible if I could grab it! \\
    14 & I could not understand the concept of the work. \\
    16 & The mechanism of music generation was unique and interesting.\\
    & The experience of chasing floating sphere and cube was also fun,\\
    & like playing in the near future, \\
    21 & I thought it would be better to experience it in a space that separated from\\ 
    & the other space so that I could feel that I was tinkering with it by myself. \\
    26 & It would be more immersive if the refresh rate and response\\
    & speed were increased.\\
    \bottomrule
  \end{tabular}
\end{table}

\begin{figure}[h]
  \centering
  \includegraphics[width=1\linewidth]{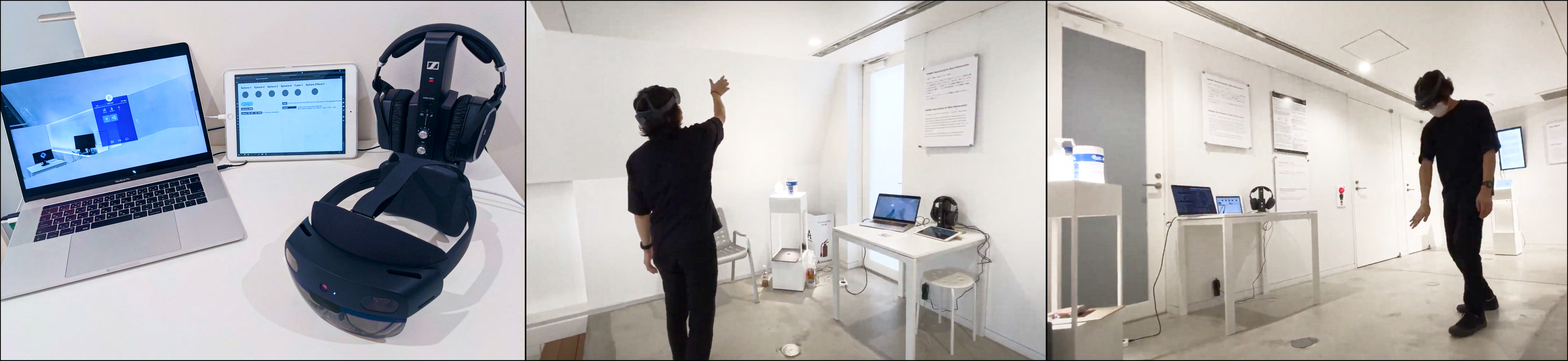}
  \caption{Installation at an art gallery in Tokyo. On a desk in the exhibition space, for the benefit of non-experienced bystanders, displays are showing a view of the visitor and the Max interface that receives the collision sound.}
  \label{installation}
\end{figure}

We created a system to experience music in the MR space, and in continuing the development, we conducted a questionnaire survey. The installation was exhibited at an exhibition in an art gallery (Figure \ref{installation}), and a 5-point (1-5) scale questionnaire was given to 27 people. The results of the responses to the questions were generally high. (Table \ref{eval-result}) In addition, we obtained optional open-ended feedback comments from participants (Table \ref{eval-comments}). There is no problem with the comfort of the operation through the basic experience and the technical novelty of this work was highly appreciated. But it is inferred from the last comment that there were issues with the instruction and care to convey the context of the artwork.

\section{Discussion}

The contribution of this work is not only to provide people with a new experience, but also to propose a new approach to listening to music and making music under the Mixed Reality context, and to suggest new ways of using AI-based music generation algorithms other than to support music composition. However, at the moment, there are still many systemic issues that prevent us from achieving the ideal interaction. The opinion in the feedback comment that it would be better to be able to grasp an object with real haptics is challenging to realize with the current system unless using additional devices that can provide tangible feedback. Refresh rate and response time are also restricted by the hardware and are difficult to improve at present. If the performance of MR devices is improved, it will be possible to provide a better, more immersive experience.

In the real world, the sounds emitted by the objects around us depend on the material and weight of the objects. For example, the sound of a collision between a virtual object and a glass on a desk will be different from that of a collision between a wall and a glass on a desk. Therefore, as a future challenge and idea, if we can grasp what kind of real objects exist around us and where they are located by running the object recognition model with the MR device camera input, we can determine the type of sound output and have variations of collision sounds. In addition, the collisions between objects in our daily life are of objects such as cups and pens, not spheres and cubes. To simplify the operation, we used primitive 3D models to make the boundary between the virtual and the real easier to understand. By preparing 3D models of everyday objects such as mugs, pens, and remote controls, as well as broken objects that we cannot throw even if we wanted to, we expect to be able to create music from interactions that are closer to the real world.

The design goal of this work should be to promote an awareness of the musical potential of the environment around us through the experience of creating music specific to a place, where the characteristics of the music produced vary according to the space. To achieve this, it is necessary to actively design exhibition spaces with different characteristics, such as brightness and abundance of objects. Ultimately, we would like to design a system that allows people to move between multiple rooms or allows whoever has HoloLens to experience the system remotely via the Internet.

\section{Conclusion}

In this study, we implemented an interface that continuously generates music from virtual object interactions with objects in real space using MR devices. By designing a new experience of assembling melodies directly from the sounds of virtual objects around us, we proposed the new application of automatic music generation technology based on machine learning and an active way of interacting with background music specific to the place. There are still technical and software design issues to be resolved, so we plan to improve it to realize better interaction and musical experience.

\small
\bibliography{main}

\begin{thebibliography}{10}

\bibitem{music_in_xr}
Luca Turchet, Rob Hamilton, and Anil Çamci.
\newblock Music in extended realities.
\newblock {\em IEEE Access}, 9:15810--15832, 2021.

\bibitem{placing_music_in_space}
Shoki Miyagawa, Yuki Koyama, Jun Kato, Masataka Goto, and Shigeo Morishima.
\newblock Placing music in space: A study on music appreciation with spatial
  mapping.
\newblock In {\em Proceedings of the 2018 ACM Conference Companion Publication
  on Designing Interactive Systems}, DIS '18 Companion, page 39–43, New York,
  NY, USA, 2018. Association for Computing Machinery.

\bibitem{spatial_music_listening_experience}
Lawrence Lim, Wei-Yee Goh, Mara Downing, and Misha Sra.
\newblock {\em A Spatial Music Listening Experience in Augmented Reality}, page
  23–25.
\newblock Association for Computing Machinery, New York, NY, USA, 2021.

\bibitem{ar_piano_impro}
Seth Glickman, Byunghwan Lee, Fu~Yen Hsiao, and Shantanu Das.
\newblock Music everywhere --- augmented reality piano improvisation learning
  system.
\newblock In {\em Proceedings of the International Conference on New Interfaces
  for Musical Expression}, pages 511--512, Copenhagen, Denmark, 2017. Aalborg
  University Copenhagen.

\bibitem{mixr}
Shalva Kohen, Carmine Elvezio, and Steven Feiner.
\newblock Mixr: A hybrid ar sheet music interface for live performance.
\newblock In {\em 2020 IEEE International Symposium on Mixed and Augmented
  Reality Adjunct (ISMAR-Adjunct)}, pages 76--77, 2020.

\bibitem{sonic_sculpture}
Charles~Patrick Martin, Zeruo Liu, Yichen Wang, Wennan He, and Henry Gardner.
\newblock Sonic sculpture: Activating engagement with head-mounted augmented
  reality.
\newblock In {\em Proceedings of the International Conference on New Interfaces
  for Musical Expression}, page 39–42. Zenodo, Jun 2020.

\bibitem{bloom_2018}
Brian Eno and Peter Shilvers.
\newblock {\em Bloom: Open Space \url{https://vimeo.com/299097246}}.
\newblock Microsoft, Nov 2018.

\bibitem{gen_music_ar_drawing}
Kyungjin Yoo and Eli Schwelling.
\newblock Spatially accurate generative music with ar drawing.
\newblock In {\em 25th ACM Symposium on Virtual Reality Software and
  Technology}, VRST '19, New York, NY, USA, 2019. Association for Computing
  Machinery.

\bibitem{komatsubara_ohta_ohashi_sonobe_nakagawa_2019}
Ryo Komatsubara, Taku Ohta, Sayuri Ohashi, Ken Sonobe, and Ryu Nakagawa.
\newblock Multimodal connector (pm), Mar 2019.

\bibitem{osc_xr}
David Johnson, Daniela Damian, and George Tzanetakis.
\newblock Osc-xr: A toolkit for extended reality immersive music interfaces.
\newblock In {\em Proc. Sound Music Comput. Conf}, pages 202--209, 2019.

\bibitem{music_vae}
Adam Roberts, Jesse Engel, Colin Raffel, Curtis Hawthorne, and Douglas Eck.
\newblock A hierarchical latent vector model for learning long-term structure
  in music.
\newblock 03 2018.

\bibitem{smart_city_sonification}
Stephen Roddy and Brian Bridges.
\newblock The design of a smart city sonification system using a conceptual
  blending and musical framework, web audio and deep learning techniques.
\newblock In {\em Proceedings of the International Conference on Auditory
  Display}, 06 2021.

\bibitem{malakai}
Zack Harris, Liam~Atticus Clarke, Pietro Gagliano, Dante Camarena, Manal
  Siddiqui, and Pablo~S. Castro.
\newblock Malakai: Music that adapts to the shape of emotions, 2021.

\bibitem{review_ai_and_xr}
Dirk Reiners, Mohammad~Reza Davahli, Waldemar Karwowski, and Carolina
  Cruz-Neira.
\newblock The combination of artificial intelligence and extended reality: A
  systematic review.
\newblock {\em Front. Virtual Real.}, 2, September 2021.

\bibitem{repo_melody_rnn}
Google Magenta.
\newblock Magenta/magenta/models/melody\_rnn at main · magenta/magenta,
  retrieved from github
  \url{https://github.com/magenta/magenta/tree/main/magenta/models/melody_rnn}.

\bibitem{repo_music_vae}
Google Magenta.
\newblock Magenta/magenta/models/music\_vae at main · magenta/magenta,
  retrieved from github
  \url{https://github.com/magenta/magenta/tree/main/magenta/models/music_vae}.

\bibitem{project_scriabin}
Chang~Min Han.
\newblock Project scriabin v.3.
\newblock In {\em Proceedings of the 7th International Conference on New
  Interfaces for Musical Expression}, NIME '07, page 388–389, New York, NY,
  USA, 2007. Association for Computing Machinery.

\bibitem{mardirossian-2007-visualizing}
Arpi Mardirossian and Elaine Chew.
\newblock Visualizing music: Tonal progressions and distributions.
\newblock In Simon Dixon, David Bainbridge, and Rainer Typke, editors, {\em
  Proceedings of the 8th International Conference on Music Information
  Retrieval, {ISMIR} 2007, Vienna, Austria, September 23-27, 2007}, pages
  189--194. Austrian Computer Society, 2007.

\end{thebibliography}
\bibliographystyle{unsrt}

\end{document}